    \documentclass[12pt,preprint]{aastex}

     \usepackage{emulateapj5}

\slugcomment{submitted to the Astrophysical Journal}

\shorttitle{Super-Eddington Luminosity Model}
\shortauthors{Kato and Hachisu}

\begin{document}


\title{A Modeling of the Super-Eddington Luminosity 
in Nova Outbursts: V1974 Cygni
  }


\author{Mariko Kato}
\affil{Department of Astronomy, Keio University, Hiyoshi, Yokohama
  223-8521 Japan  }
\email{mariko@educ.cc.keio.ac.jp}

\and

\author{Izumi Hachisu}
\affil{Department of Earth Science and Astronomy, College of Arts and 
Sciences, University of Tokyo, Komaba, Meguro-ku, Tokyo 153-8902, Japan}
\email{hachisu@chianti.c.u-tokyo.ac.jp}


\begin{abstract}
We have modeled nova light curves exceeding the Eddington luminosity.
It has been suggested that a porous structure develops in nova
envelopes during the super Eddington phase and the effective opacity
is much reduced for such a porous atmosphere.
Based on this reduced opacity model, we have calculated envelope
structures and light curves of novae.  The optically thick wind model
is used to simulate nova winds.
We find that the photospheric luminosity and the wind mass-loss rate
increase inversely proportional to the reducing factor of opacities, 
but the wind velocity hardly changes.  We also reproduce the optical
light curve of V1974 Cygni (Nova Cygni 1992)
in the super-Eddington phase, which lasts 13 days from the optical
peak 1.7 mag above the Eddington luminosity.
\end{abstract}




\section{Introduction}

The super-Eddington luminosity is one of the long standing problems
in nova theory.  The peak luminosity of classical novae often exceeds 
the Eddington limit by a factor of several 
\citep[][and references therein]{del95}.
Super-Eddington phases last several days or more,
longer than the dynamical time scale of white dwarf (WD) envelopes. 
Many theoretical works have attempted to reproduce this phenomena, but  
not succeeded yet
\citep[e.g.,][]{pri78, spa78, nar80, sta85, pri86, kut89, pri92, kov98}.

The Eddington luminosity 
is understood as an upper limit of the luminosity of stars in  
hydrostatic equilibrium.  Its classical definition is 

\begin{equation}
L_{\rm Edd,cl} = {4\pi cGM \over\kappa_{\rm el}},
\end{equation}

\noindent
where $c$ is the light speed, $G$ the gravitational constant,
$M$ the mass of the WD, and $\kappa_{\rm el}$ is the opacity
by electron scattering.  If the diffusive luminosity exceeds
this limit, the stellar envelope cannot
be in hydrostatic balance and a part of the envelope is ejected.

During nova outbursts nuclear burning produces energy much faster
than this limit.  The envelope is accelerated deep inside 
the photosphere and a part of it is ejected as a wind.
Once the wind occurs, the diffusive luminosity is consumed to drive
the wind.  As a result, the photospheric luminosity decreases 
below the Eddington limit defined by equation 
(1) \citep{kat83, kat85}.

Recently, \citet{sha01b,sha02} presented a new idea of clumpy atmospheres 
to explain the super-Eddington luminosity of novae. 
The nova envelope becomes unstable against clumpiness shortly after 
the ignition when the luminosity exceeds a critical fraction 
of the Eddington limit \citep{sha01a}. Such a clumpy structure 
reduces effective opacity, and correspondingly, 
increases the effective Eddington luminosity. Therefore, the luminosity 
could be larger than the classical Eddington limit, 
even though it does still not exceed the effective Eddington luminosity.

\citet{sha01b} suggested a model of nova envelope with the super-Eddington 
luminosity  consisting of four parts: (1) convective region: 
a bottom region of the envelope in which the diffusive luminosity 
is sub-Eddington and additional 
energy is carried by convection, (2) a porous atmosphere:
the effective Eddington luminosity is larger than the classical Eddington 
limit, (3) an optically thick wind region: the effective Eddington 
limit tends to the classical value, and (4) the photosphere and above.
 
Based on Shaviv's picture, we have assumed reduced opacities  
to model the super-Eddington phase of V1974 Cyg (Nova Cygni 1992). 
V1974 Cyg  is a well observed classical nova so that various 
multiwavelength observations are available such as optical 
\citep{iij03}, supersoft X-ray \citep{kra02}, and radio \citep{eyr05}.
\citet{cho97} summarized observational estimates of optical maximum 
magnitude, ranging from  $-7.3$ to  $-8.3$ mag with an average magnitude 
of $-7.78$.  These values indicate that the peak luminosity exceeded
the Eddington limit by more than a magnitude and the duration of
super-Eddington phase lasts several days or more. 

In \S 2, we briefly describe our numerical method. Physical properties 
of the envelope with reduced effective opacities are shown in \S 3. Our 
light curve model of V1974 Cyg is given in \S 4.
Discussion follows in \S 5.


\section{Envelope model with reduced opacity}

We have calculated structures of envelopes on mass-accreting WDs 
by solving the equations of motion, mass continuity, energy generation,
and energy transfer by diffusion.  The computational method and
boundary conditions are the same as those in \citet{kat94} except
the opacity. We use an arbitrarily reduced opacity

\begin{equation}
\kappa_{\rm eff} = \kappa/s,
\end{equation}

\noindent
where $\kappa$ is OPAL opacity \citep{igl96} and $s$ is a opacity 
reduction factor that represents reduced ratio of opacity due to 
clumpiness of the envelope.
The effective Eddington luminosity now becomes

\begin{equation}
L_{\rm Edd,eff} = {4\pi cGM\over{\kappa_{\rm eff}}}.
\end{equation}

\noindent
When $s$ is greater than unity, the luminosity can be larger than 
the classical Eddington limit (1).
Note that the Eddington luminosity (3) is a local variable because
OPAL opacity is a function of local variables. 

As a first step, we simply assume that the opacity reduction factor $s$  
is spatially constant.
Figure 1 shows numerical results for 
three envelopes of $s=1$, 3, and 10 on a $1.0 ~M_\sun$ WD with the 
Chandrasekhar radius,  i.e., $\log R_{\rm WD} ~{\rm (cm)}=8.733$.
The chemical composition of the envelope is assumed to be uniform,
i.e., $X=0.35$, $Y=0.33$, $C+O$=0.3, and $Z=0.02$, where $Z$ includes
carbon and oxygen by solar composition ratio for heavy elements.

In Figure 1, the effective Eddington luminosity (3) is plotted by 
dashed lines, which sharply decreases at $\log r ~{\rm (cm)} \sim$ 11.1
corresponding to the iron peak in OPAL opacity 
at $\log T ~({\rm K}) ~\sim 5.2$.
The wind is accelerated in this region and 
reaches a terminal velocity deep inside the photosphere. 
The diffusive luminosity ($L_{\rm r}$) decreases outward because
the energy flux is consumed to push matter up against the gravity. 
These features are qualitatively the same as those in the three nova envelopes
with $s=1,$ 3, and 10.

\begin{figure}
\plotone{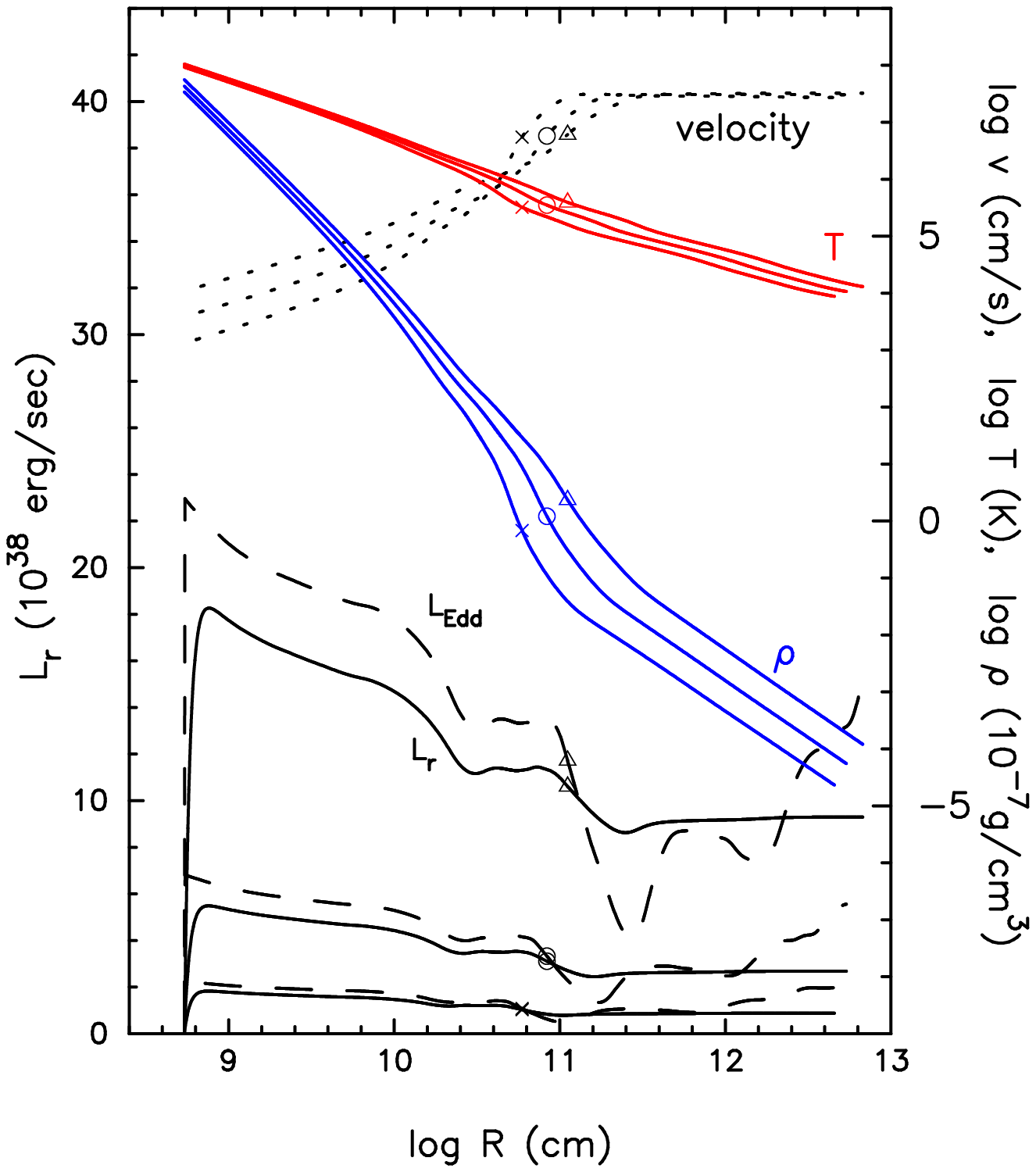} 
\caption{Velocity ({\it dotted}), temperature ($T$: {\it solid}), 
density ($\rho$: {\it solid}), local Eddington luminosity 
($L_{\rm Edd}$: {\it dashed}), and diffusive luminosity 
($L_{\rm r}$: {\it solid}) for three envelope solutions
on a $1.0 M_{\odot}$ WD.
The critical points of steady state winds are denoted by crosses, 
circles, and triangles for three solutions with $s$=1, 3, and 10, 
respectively. 
The right edge of each line corresponds to the photosphere.
 \label{fig1}}
\end{figure}

\begin{figure}
\plotone{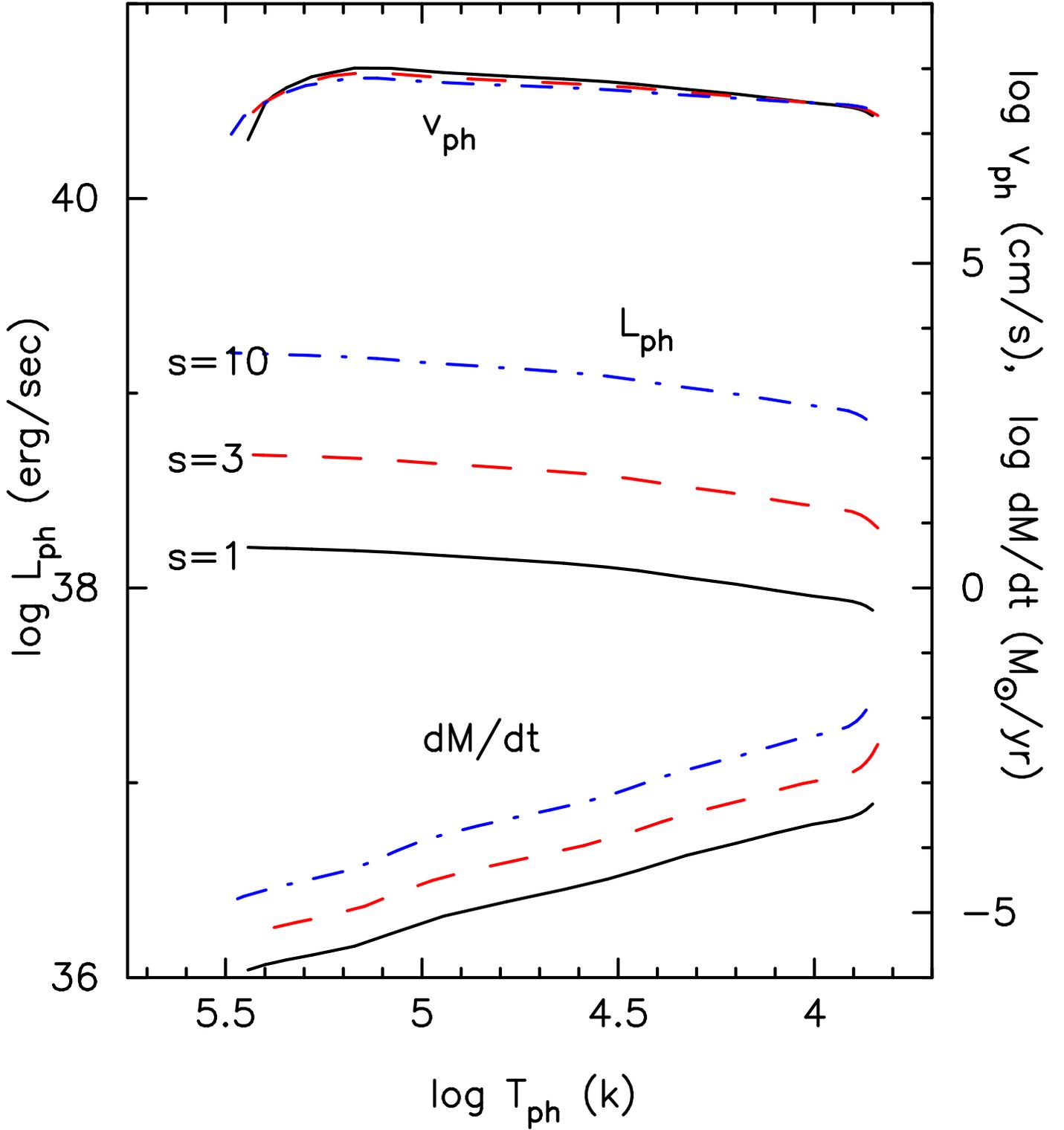} 
\caption{ Photospheric velocity ($v_{\rm ph}$), photospheric luminosity 
($L_{\rm ph}$), and wind mass loss rate ($dM/dt$) for three evolutionary
sequences, $s$=1 ({\it solid}), 3 ({\it dashed}), 
and 10 ({\it dash-dotted}) against the photospheric temperature 
($T_{\rm ph}$).  The value of $s$ is attached to each curve.  
Novae evolve from low to high photospheric temperatures.
 \label{fig2}}
\end{figure}

\placefigure{fig1}
\placefigure{fig2}

Figure 2 shows the photospheric velocity ($v_{\rm ph}$), the wind mass 
loss rate ($\dot M$), and the photospheric 
luminosity ($L_{\rm ph}$)  for three evolutionary sequences
of $s=1, 3,$ and 10.  The $s=1$ sequence is already reported
in \citet{kat94}.  In each evolutionary sequence, 
the envelope mass is large for smaller photospheric
temperature ($T_{\rm ph}$).  The figure also shows that 
$L_{\rm ph}$ and $\dot M$ increase almost proportionally to
$s$, whereas the wind velocity ($v_{\rm ph}$) hardly changes but
even slightly decreases. 

Theoretical light curves are calculated from these sequences. 
After the onset of a nova outburst, the envelope expands 
to a giant size and the luminosity reaches its peak. After that, 
the envelope mass gradually decreases owing mainly
to the wind mass loss.  During the nova decay phase, the bolometric 
luminosity is almost constant whereas the photospheric temperature 
increases with time.  The main emitting wavelength region moves
from optical to supersoft X-ray through ultra-violet (UV). 
Therefore, we obtain decreasing visual magnitudes \citep{kat94}. 

\begin{figure}
\plotone{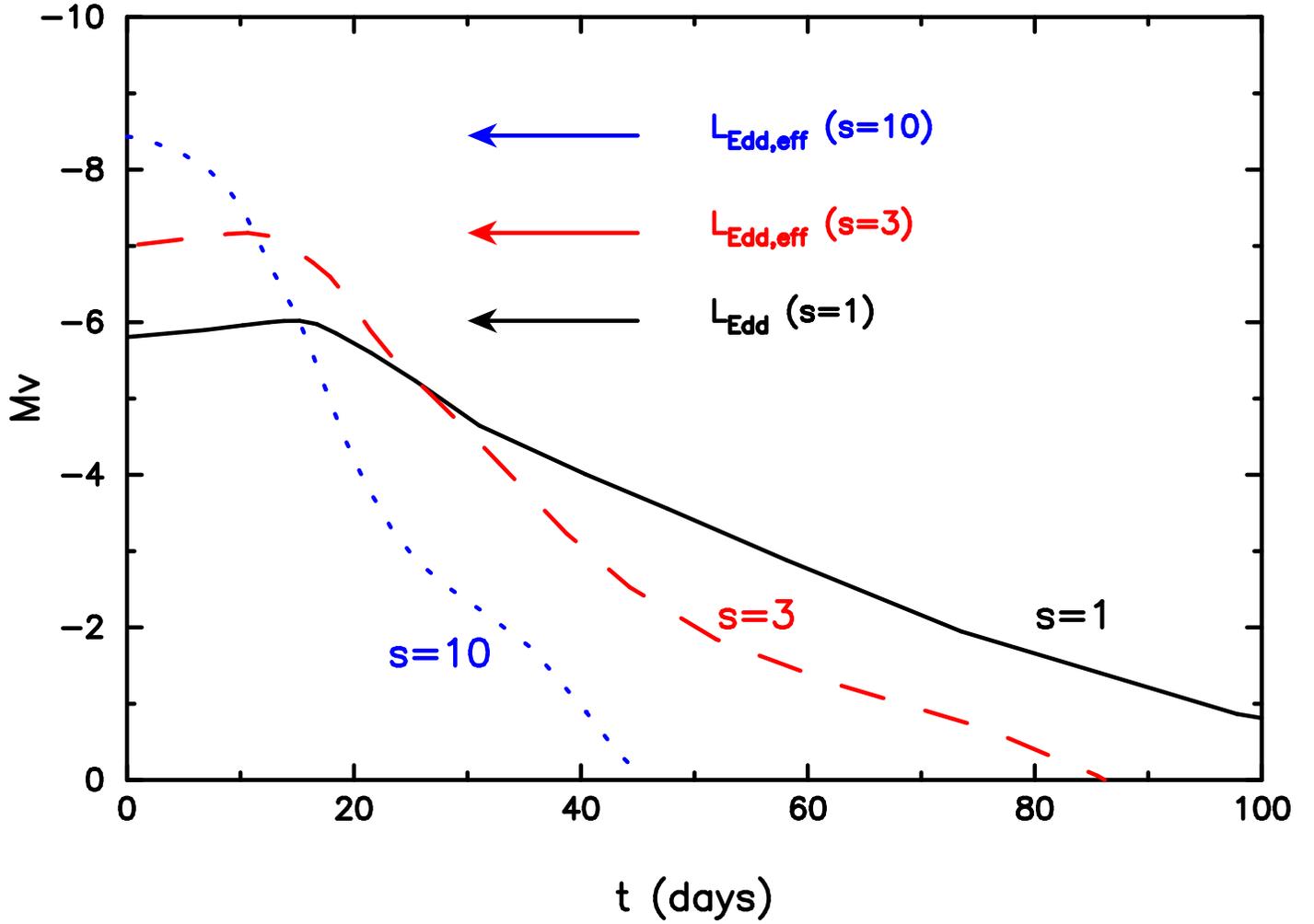} 
\caption{Theoretical light curves of a $1.0 ~M_{\odot}$ WD
for reduced opacities of $s=1$, 3, and 10.
The value of $s$ is attached to each curve.
The maximum visual magnitude of each light curve is indicated
by an arrow.  
 \label{fig3}}
\end{figure}

\placefigure{fig3}

Figure 3 shows visual light curves for the opacity reduction 
factor $s=1$, 3, and 10.  The visual magnitude decays quickly
for a larger $s$ because an envelope for a larger $s$ has 
a heavier wind mass loss. 
The peak luminosity of each light curve is shown by arrows.
When the opacity reduction factor $s$ is larger than unity, 
the peak luminosity exceeds the classical Eddington limit,
which is roughly corresponding to the Eddington luminosity 
for $s=1$.

\section{Light Curve of V1974 Cyg}

Recently, \citet{hac05} presented a light curve model of V1974 Cyg 
that reproduced well the observed X-ray, UV, and optical light 
curves except for a very early phase 
of the super-Eddington luminosity.  
Here, we focus on this early phase ($m_{\rm v} \geq 6$) and 
reproduce the super-Eddington luminosity based on 
the reduced opacity model.

We adopt various WD model parameters after their best fit model,
i.e., the mass of $1.05 ~M_\sun$, radius of $\log ~(R/R_\sun)=-2.145$,
and chemical composition of $X$=0.46, $CNO=0.15$, $Ne=0.05$,
and $Z=0.02$ by mass.  These parameters are determined from 
the X-ray turn-off time, epoch at the peak of UV 1455 \AA~ flux,
and epoch at the wind termination.  All of these epochs are
in the post super-Eddington phase.

Our simple model with a constant $s$ such as in Figure 3
does not reproduce the observed light curve of V1974 Cyg.
Therefore, we assumed that $s$ is a decreasing function of time.
Here, the decreasing rate of $s$ is determined from the wind 
mass loss rate and the envelope mass of solutions we have chosen. 
After many trials, we have found that we cannot obtain a light curve 
as steep as that of V1974 Cyg. 
Finally, we further assume that $s$ is a function both of
temperature and time.  We define $s$ as  unity 
in the outer part of the envelopes ($\log T < 4.7$), 
but a certain constant value ($s > 1$) in the inner region 
($\log T > 5.0$), 
and changes linearly between them. This assumption well represents 
the nova envelope model by \citet{sha01b} outlined in \S1.
After many trials, we choose $s=5.5$ at the optical
peak (JD 2,448,676) and gradually decreases it to 1.0 with time  
as shown in Figure 4.
The choice of $s$ is not unique; we can reproduce visual light curve by
adopting another vale of $s$. Here, we choose $s$ to reproduce not only 
$V$ band magnitudes but also  UV 1455 \AA~ continuum 
fluxes \citep{cas04}. This is a strong constraint for a choice of $s$, 
and thus, we hardly find another $s$ that reproduce both 
visual and UV light curves.

\begin{figure}
\plotone{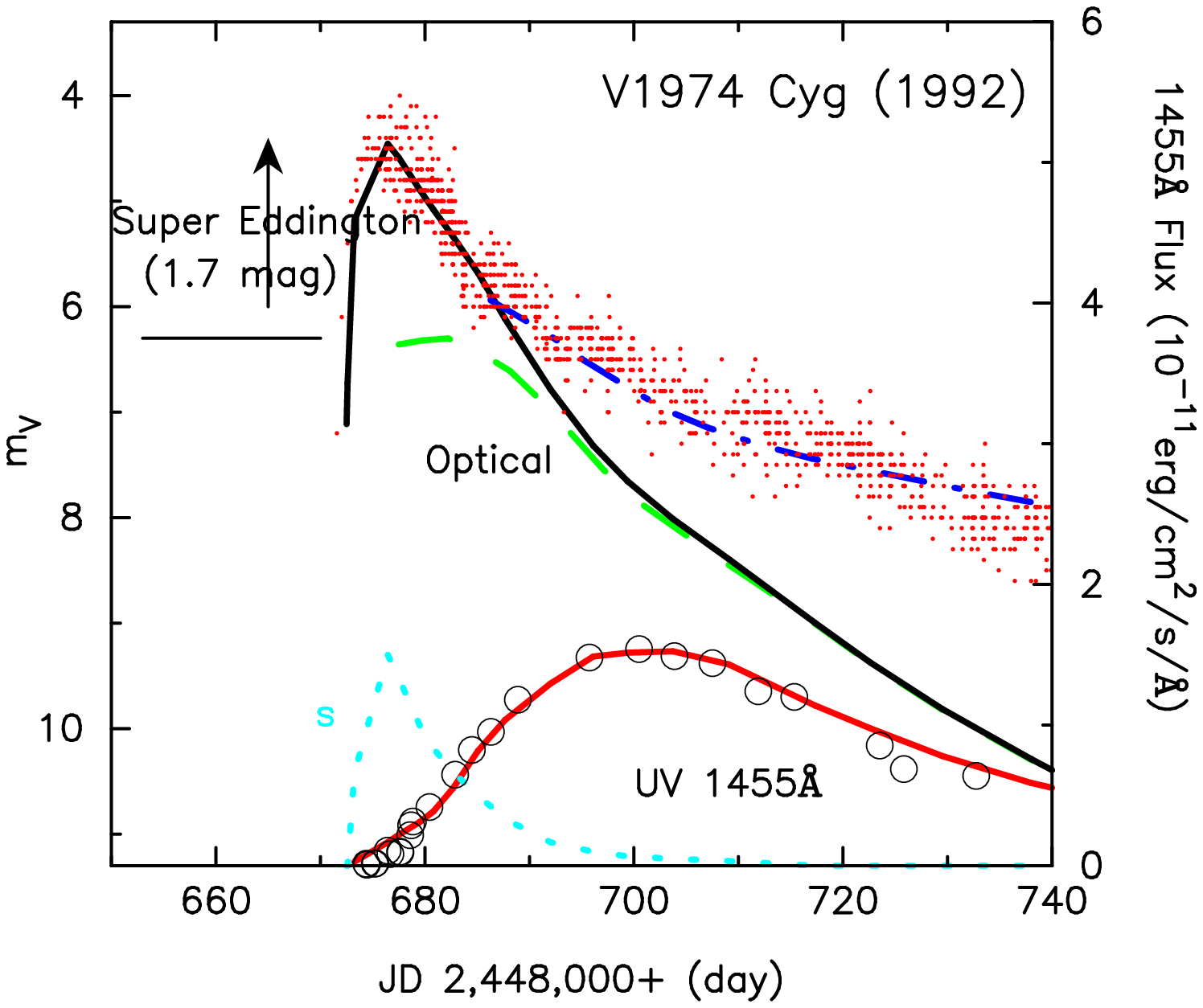}
\caption{Theoretical light curves of our $1.05 ~M_\sun$ WD model
for a variable $s$ model: 
visual ({\it upper solid line}) and UV 1455 \AA~ continuum 
({\it lower solid line}).
Observational visual magnitudes of V1974 Cygni ({\it dot}) are 
taken from AAVSO and observational UV ({\it circle}) are
from \citet{cas04}.
{\it Dashed}: light curve of $1.05 ~M_\sun$ WD for $s=1$.
{\it Dash-dotted}:  visual flux from free-free emission
model by \citet{hac05}.
{\it Dotted}: $s$ is shown in a linear scale between 1.0
(at the bottom) and 5.5 (at the peak). 
 \label{fig4}}
\end{figure}

\placefigure{fig4}

Figure 4 depicts our modeled light curve that reproduces 
well both the early optical and UV 1455 \AA~  continuum light curves. 
The observed UV flux is small even 
in the super-Eddington phase in which the photospheric luminosity is several 
times larger than that in the later phase. This means that the photospheric
temperature is as low as $\log T < 4.0$. In our model, the temperature is 
$\log T = 3.93$ at the optical peak and lower than 
4.0 for 8 days after the peak, gradually increasing with time.  
Such a behavior 
is consistent with $B-V$ evolution reported by \citet{cho93}, in which  
$B-V$ is larger than 0.3 for the first ten days from JD 2,448,677 and 
gradually decreases with time. 

In the later phase, our modeled visual magnitude decays too 
quickly and is not compatible with the 
observed data. \citet{hac05} concluded that this excess comes
from free-free emission from optically thin plasma outside the photosphere.
They reproduced well the optical light curve in the late phase
by free-free emission as shown by the dash-dotted line in Figure 4.

We see that the peak luminosity exceeds the Eddington limit by 1.7 mag, and 
the super-Eddington phase lasts 12 days after its peak.

The distance to the star is obtained from the comparison between observed and 
calculated UV fluxes, that is, 1.83 kpc with $A_{\lambda}=8.3 E(B-V)=2.65$
 for $\lambda$=1455 \AA~ \citep{sea79}.  
From the comparison of optical peaks, the distance is also obtained to be 
1.83 kpc with $A_V$=0.99 \citep{cho97}. 
This value is consistent with the distance discussed by \citet{cho97}
that ranges from 1.3 to 3.5 kpc with a most probable value of 1.8 kpc
\citep[see also][]{ros96}.

\section{Discussion}

\citet{sha01b} found two types of radiation-hydrodynamic instabilities  
in plane parallel envelopes. 
The first one takes place when $\beta$ decreases 
from 1.0 (before ignition) to $\sim 0.5$ 
and the second one occurs when $\beta$ decreases to $\sim 0.1$. 
Here $\beta$ is the gas pressure divided by the total pressure. 
When the luminosity increases to a certain value 
the envelope structure changes to a porous one
in a dynamical time scale.  Radiation selectively goes through
relatively low-density regions of a porous envelope.
\citet{sha98} estimated effective opacities in inhomogeneous 
atmospheres and showed that they always less than the original opacity for 
electron scattering, but can be greater than the original one in some cases
 of Kramer's opacity.  
 \citet{rus05} have calculated radiative transfer 
in slab-like-porous atmospheres, and found 
that the diffusive luminosity is about 5-10 times greater than the classical 
Eddington luminosity when the density ratio of
porous structures is higher than 100.
  
  In nova envelopes, we do not know either how clumpy structures
develop to reduce the effective opacity or how long such porous
density structures last.  The exact value of the opacity reduction
factor $s$ is uncertain until time-dependent non-linear calculations
for expanding nova envelope will clarify the typical size and
the density contrast in clumpy structures.  Therefore, in the present
paper, we have simply assumed that $s$ is a function of temperature
and made it to satisfy the condition that $s$ is larger than unity 
deep inside the envelope and approaches unity near the photosphere.

The anonymous referee has suggested that $s$ may be a function of
``optical width'' over a considering local layer rather than
a function of temperature.  Here, the ``optical width'' means
the optical length for photons to across the local clumpy layer
in the radial direction.  If this ``optical width'' is smaller
than unity or smaller than some critical value,
the porous structure hardly
develops and then we have $s=1$.  In the opposite case,
the porous structure develops to reduce the effective opacity
and then we have $s$ much larger than unity.
The ``optical width'' description may be a better expression 
for the opacity reduction factor $s$,  because the relation between the opacity reduction
factor and porous structure is clearer.

  We have estimated the ``optical width'' ($\delta \tau$)
of a local layer using 
the solution at the optical peak in Figure 4: it is $\delta \tau \sim 3$
near the photosphere, 19 at $\log T=4.76$, 580 at $\log T=5.56,
2.8\times 10^4$  at $\log T=6.36$, and $2\times 10^7$  at $\log T=8.03$,
i.e., the nuclear burning region.
Here, we assume that the ``geometrical width'' of the local layer
is equal to the pressure scale hight, $r/(d\ln P/d\ln r)$.
This ``optical width'' decreases quickly outward and reach
the order of unity in the surface region, i.e., the ``optical width'' is large at
high temperature regions and small in low temperature regions. Therefore, 
we regard that our
assumption of $s$ qualitatively represents the dependence of
the opacity reduction factor on the ``optical width''of a local layer.

  In our computational method, this ``optical width'' is calculated
only after a solution is obtained after many iterations to adjust
boundary conditions.  The feedback from the ``optical width'''
requires further huge iterations.  Therefore, in the present paper,
we assume a simple form of $s$.

 The wind acceleration is closely related to spatial change of the
effective opacity. In a case of varying $s$, for example, 
when it is a function of temperature, $s$ determines the wind
acceleration.  If we assume the other spatial form of $s$, 
the acceleration is possibly very different.  In our case
in Figure 4, $s$ is a monotonic function, and then the reduced opacity 
still has a strong peak at $\log T \sim 5.2$ although the peak value is 
smaller by a factor of $s$ than that of the OPAL peak.
The resultant velocity profile is essentially the same as
those in Figure 1; the wind is accelerated  at the shoulder of
the OPAL peak.

\acknowledgments
We thank A. Cassatella for providing us with their machine readable 
UV 1455 \AA~ data of V1974 Cyg and also AAVSO for the visual 
data of V1974 Cyg.   We thank an anonymous referee 
for useful and valuable comments that improved the manuscript.
This research was supported in part by the 
Grant-in-Aid for Scientific Research
(16540211, 16540219) of the Japan Society for the Promotion of Science.

\end{document}